\documentclass{article}
\usepackage{spconf,amsmath,graphicx}
\usepackage{cite}
\usepackage{amsmath,amssymb,amsfonts}
\usepackage{algorithmic}
\usepackage{graphicx}
\usepackage{textcomp}

\usepackage{siunitx}
\usepackage{bm}
\usepackage{multirow}
\usepackage[caption=false, font=footnotesize, position=top]{subfig}

\usepackage{hyphenat}

\newcommand{\etal}{\textit{et~al.} }

\usepackage[firstpage=true]{background}
\usepackage{hyperref}

\SetBgContents{%
	\fbox{%
		\parbox{\textwidth}{%
			\footnotesize © 2023 IEEE. Personal use of this material is permitted. Permission from IEEE must be
obtained for all other uses, in any current or future media, including
reprinting/republishing this material for advertising or promotional purposes, creating new
collective works, for resale or redistribution to servers or lists, or reuse of any copyrighted
component of this work in other works. DOI: \href{https://doi.org/10.1109/ACCESS.2023.3323873}{10.1109/ACCESS.2023.3323873.}}}%
}	
\SetBgScale{1}
\SetBgAngle{0}
\SetBgPosition{current page.north}
\SetBgVshift{-1.6cm}
\SetBgColor{black}
\SetBgOpacity{1}

\title{Learned Wavelet Video Coding using Motion Compensated Temporal Filtering}
\name{Anna Meyer, Fabian Brand, and Andr\'e Kaup\thanks{The authors gratefully acknowledge that this work has been funded by the Deutsche Forschungsgemeinschaft (DFG, German Research Foundation) under project number 461649014.}}
\address{\textit{Multimedia Communications and Signal Processing} \\
	\textit{Friedrich-Alexander-Universität Erlangen-Nürnberg (FAU)}\\
	Erlangen, Germany \\
	\{ anna.meyer, fabian.brand, andre.kaup \}@fau.de}

\def\BibTeX{{\rm B\kern-.05em{\sc i\kern-.025em b}\kern-.08em
		T\kern-.1667em\lower.7ex\hbox{E}\kern-.125emX}}
\begin{document}
	
	\maketitle
	
\begin{abstract}
	This paper presents an end-to-end trainable wavelet video coder based on motion-compensated temporal filtering. Thereby, it introduces a different coding scheme for learned video compression, which is dominated by residual and conditional coding approaches. By performing discrete wavelet transforms in temporal, horizontal, and vertical dimensions, an explainable framework with spatial and temporal scalability is obtained. This paper investigates a novel trainable motion-compensated temporal filtering module implemented using the lifting scheme. It demonstrates how multiple temporal decomposition levels can be considered during training. Furthermore, larger temporal displacements owing to the coding order are addressed and an extension adapting to different motion strengths during inference is introduced. The experimental analysis compares the proposed approach to learning-based coders and traditional hybrid video coding. Especially at high rates, the approach exhibits promising rate-distortion performance. The proposed method achieves average Bj{\o}ntegaard Delta savings of up to 21\% over HEVC, and outperforms state-of-the-art learned video coders.
\end{abstract}


\section{Introduction}
\label{sec:intro}

Following the progress of learned image compression, there have been significant advances in learned video compression. Built on learned image coders, video coding approaches exploit temporal redundancies by following two main paradigms: residual and conditional coding. Residual coders \cite{Lu2019CVPR, Lu2021,Yang2021, Hu2021CVPR, Hu2022CVPR, Agustsson2020, Rippel2021, Park2021, Yang2023, Thang2023} largely take over the structure of known hybrid video coders such as VVC \cite{Bross2021}. Using motion-compensated inter prediction, the residual between the predicted and current frame is compressed and then transmitted. Instead of transmitting a difference signal, conditional coders compress the current frame directly under the condition that both the encoder and decoder know the prediction. Since the introduction of conditional coding in learned video compression by a framework called DCVC \cite{li2021deep}, there have been several improvements of DCVC \cite{Sheng2022, Li2022} as well as other conditional coding schemes based on a generative model \cite{Ho2022} or transformers \cite{Mentzer2022, xiang2023mimt}. With these developments, conditional coding currently outperforms residual coders and represents the state of the art in learned video coding.

\begin{figure}[tb]
	\centering
	\includegraphics[width=0.48\textwidth]{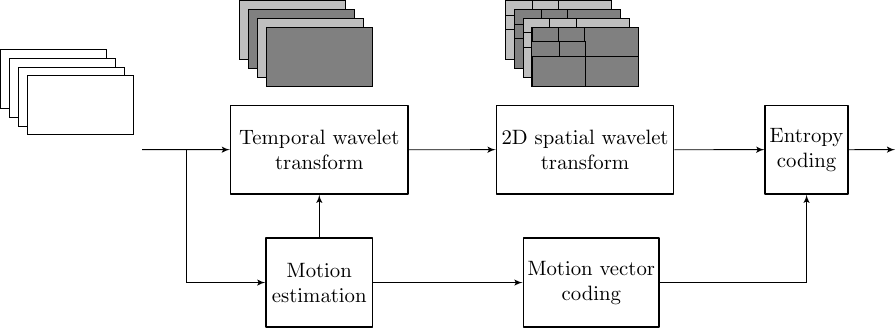}
	\caption{Schematic overview of the wavelet video coding scheme. This paper introduces a novel trainable version of the coding scheme. A temporal wavelet transform is followed by a 2D wavelet transform in horizontal and vertical dimensions. By incorporating motion compensation into the temporal wavelet transform, Motion-Compensated Temporal Filtering (MCTF) is performed. }
	\label{fig:mctf}
\end{figure}

This paper investigates a different coding scheme visualized in Fig.~\ref{fig:mctf}: learned wavelet video coding. It performs a discrete wavelet transform (DWT) in temporal, horizontal, and vertical dimensions. Specifically, an end-to-end trainable wavelet video coder based on Motion-Compensated Temporal Filtering (MCTF) \cite{Ohm1994} is introduced. Traditional MCTF as proposed by Ohm \cite{Ohm1994} and improved by Choi and Woods \cite{Choi1999}, incorporates motion compensation into the temporal wavelet transform. Until the early 2000s, MCTF-based wavelet video coding was an active research topic as a scalable alternative to predictive transform coders. With the success of the video coding standard H.264/AVC \cite{Wiegand2003}, hybrid video coding approaches have dominated the field. 
Transform coding has emerged as the predominant principle in learned image and video compression. Here, the foundation of most popular coders \cite{Balle2016, balle2018variational, minnen2018joint} is based on nonlinear transform coding \cite{Balle2021}. 

Recently, employing a learned spatial wavelet transform for end-to-end image compression has shown great potential by achieving state-of-the-art performance \cite{maiwave++}. Motivated by this emerging topic of trainable wavelet transforms for compression, the novel learned MCTF video coding approach is built on top of the wavelet image coder called iWave++~\cite{maiwave++}. The MCTF video coder provides a flexible framework that supports lossless compression. In addition, MCTF enables a fully scalable video coder. Compared to other learned approaches, which usually do not support input in YUV 4:2:0 format, wavelet video coding allows arbitrary input formats.

The focus of this paper is on the investigating a novel trainable MCTF module and compressing the obtained temporal subbands with the state-of-the-art wavelet image coder iWave++ \cite{maiwave++}. The contributions of this paper are as follows:

\begin{itemize}
	\item Introduction of the first end-to-end trainable wavelet video coding scheme. To date, there have been no learned video compression approaches based on MCTF.
	\item Presentation of a training strategy for multiple temporal decomposition levels in MCTF.
	\item Investigation of large temporal displacements due to the MCTF coding structure and a first solution for handling these cases more efficiently.
	\item Proposal of a content-adaptive MCTF approach that adapts to different types of motion during inference.
\end{itemize}

\section{State of the Art} 
\noindent
\subsection{Learned Video Compression} 
\noindent
DVC \cite{Lu2019CVPR, Lu2021} was the first learning-based deep video compression framework. It follows the structure of a traditional hybrid P-frame codec but replaces its modules for motion estimation, motion vector and residual compression by neural networks. With the feature-space video coding network FVC \cite{Hu2021CVPR}, the DVC framework was significantly improved by performing these operations in the feature space. The coarse-to-fine framework C2F \cite{Hu2022CVPR} further advanced residual coding using two-stage motion compensation at different resolutions and mode prediction networks. 

Conditional coding can offer theoretical benefits over residual coding \cite{Brand2022a} and learning-based frameworks allow for its straightforward implementation via conditional autoencoders \cite{Brand2021, Ladune2020}. The DCVC \cite{li2021deep} framework has attracted greater attention to conditional coding for learned video compression. Conditioning on the temporal context in the feature space \cite{Sheng2022}, and an extended entropy model with a latent prior in addition to quantization at different granularities \cite{Li2022} made the DCVC framework reach state-of-the-art performance. Another conditional coding approach \cite{xiang2023mimt} follows the structure of DCVC but uses a transformer-based entropy model. There are also frameworks based on augmented normalizing flows \cite{Ho2022} or without an explicit motion model, such as the video compression transformer VCT \cite{Mentzer2022}.
 \subsection{Wavelets for Learned Image Compression} 
\begin{figure}[tb]
	\centering
	\includegraphics[width=0.475\textwidth]{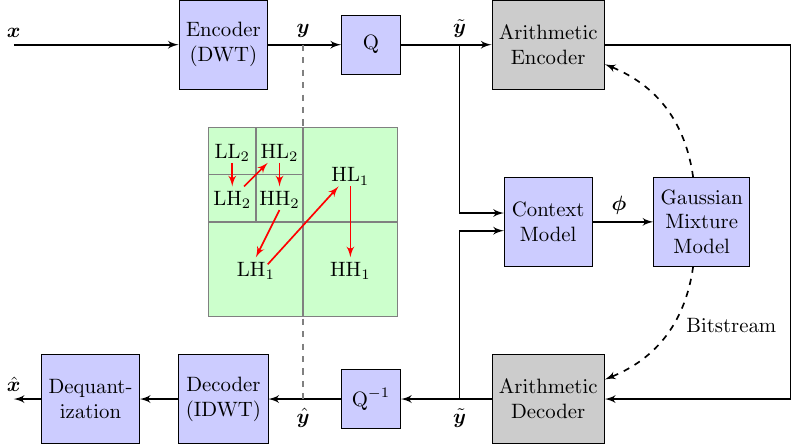}
	\caption{Overview of the end-to-end image compression method iWave++ \cite{maiwave++}. $x$ is a single luma or chroma channel of an image in the YCbCr color space. The red arrows indicate the coding order of the subbands $y$. Trainable modules are colored in blue. For visualization, the subbands of two decomposition levels are shown. }
	\label{fig:iwave++}
\end{figure}
\noindent
The traditional discrete wavelet transform has desirable properties for image and video coding. Its compromise between spatial and frequency resolution fits the correlation structure of image data: edges can be coded more efficiently in the spatial domain, whereas smooth shades and regular textures can be better modeled in the frequency domain. Hence, the image compression standard JPEG2000~\cite{Taubman2002} and the Dirac video coder~\cite{dirac} employ a DWT as an alternative to the discrete cosine transform. 
The coders rely on the lifting scheme \cite{sweldens1995NewPhil} for a fast and efficient implementation of the DWT. With the lifting structure, the DWT can be performed in place by factoring its calculation into multiple lifting steps. At the same time, the lifting structure allows the construction of new wavelet filters, so-called second-generation wavelets.  Moreover, the lifting scheme is a reversible structure and is thus well suited for realizing lossless transforms that can be incorporated into learning-based frameworks.

Convolutional Neural Networks (CNNs) allow the optimization of wavelet transforms based on a set of training images~\cite{Ma2020}. Such a learned wavelet transform implemented via the lifting scheme has been shown to outperform the wavelet filters used by JPEG2000. The learned wavelet transform forms the basis of the end-to-end trainable wavelet image coder iWave++~\cite{maiwave++}. An overview of iWave++ is shown in Fig.~\ref{fig:iwave++}.  First, the encoder performs a CNN-based DWT with four decomposition levels. The obtained tree-structured subbands constitute a hierarchical representation at different resolutions. For transmission, the wavelet coefficients are quantized using scalar quantization with a trainable parameter. Subsequently, a CNN-based context model estimates the entropy parameters of a Gaussian mixture model employed for adaptive arithmetic coding. The context model exploits correlations within the current subband to be coded and across subbands from different decomposition levels. After an inverse discrete wavelet transform (IDWT) is performed by the decoder, a post-processing module compensates for quantization artifacts. 

Learned wavelet image compression provides a flexible framework. A 3D version of iWave++ \cite{Xue2021} has been employed for lossless and lossy medical image compression, that is, for coding 3D volume data without temporal information. An extension through an affine wavelet transform module further improved volumetric image compression performance \cite{Xue2023}. The low- and highpass subbands obtained from the lifting scheme are re-scaled by an affine map computed based on the output of the prediction and update filters.

Dong \etal \cite{Dong2022} proposed a partly trainable wavelet video coder that follows a ''t+2D'' decomposition structure. First, they perform a temporal wavelet transform taken from \cite{Liu2017}. Afterwards, they code the obtained temporal subbands using a trainable entropy parameter estimation module that largely takes over the structure of iWave++. In addition, Dong \etal enabled quality scalability via bitplane coding. 
This paper focuses on a trainable temporal wavelet transform instead to obtain a fully CNN-based wavelet video coder.

\begin{figure}[tb]
	\centering	
	\includegraphics[width=0.49\textwidth]{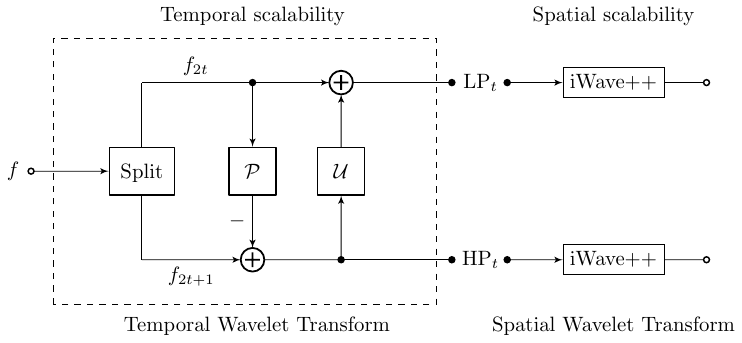}	  
	
	\caption{Overview of the proposed wavelet video coding scheme for one temporal decomposition level with two frames. $\bm{f}$ denotes the input video sequence.}
	\label{fig:overview}
\end{figure}
\section{Learned Wavelet Video Coding} \label{sec:iwave++}
\noindent
In the following section, the end-to-end trainable wavelet video coding scheme is introduced. Fig.~\ref{fig:overview} provides an overview of the proposed approach. The temporal wavelet transform realized via MCTF provides temporal scalability. The obtained temporal low- and highpass subbands are coded using dedicated iWave++ \cite{maiwave++} image compression models. Its spatial 2D wavelet transform yields spatial scalability. 

First, the concept of wavelet video coding for one temporal decomposition level, that is, for coding two frames is explained. Subsequently, multiple temporal decomposition levels are discussed in Section~\ref{sec:multiple}.

\subsection{Trainable Temporal Wavelet Transform}
\noindent \vspace{-1cm}
\subsubsection{Lifting Scheme} 
The lifting structure \cite{sweldens1995NewPhil} provides a flexible and efficient implementation of the DWT.
The temporal lifting scheme illustrated in Fig.~\ref{fig:overview} consists of the three steps split, predict, and update. In the first step, the input video sequence $f$ is split into even- and odd-indexed frames $f_{2t}$ and $f_{2t+1}$. In the next step, the odd frames are predicted from the even frames with the prediction operator $\mathcal{P}$. A temporal highpass subband (HP$_t$) is obtained as $\mathrm{HP}_t = f_{2t+1} - \mathcal{P}(f_{2t})$. Subsequently, an update step is performed according to $\mathrm{LP}_t = f_{2t} + \mathcal{U}(\mathrm{HP}_t)$ resulting in a temporal lowpass subband (LP$_t$). The inverse lifting scheme is obtained by reversing the order of the operations and inverting the signs. Rounding the output of the prediction and update operators yields an integer-to-integer temporal DWT required for lossless reconstruction \cite{Calderbanka}.

\subsubsection{Prediction and Update Filters}
\begin{figure}[tb]
	\centering	
	\subfloat{%
		\includegraphics[width=0.48\textwidth]{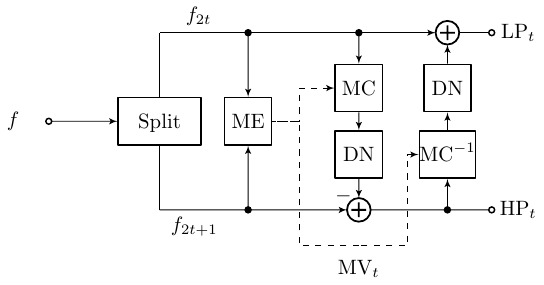}	  
	}
	\caption{Details on prediction and update filters. $\bm{f}$ denotes the input video sequence. The ''ME'' module contains motion estimation and motion vector coding. Its output MV$_t$ corresponds to the decoded motion vectors at time instance $t$. MC stands for motion compensation and MC$^{-1}$ for inverse motion compensation. The ''DN'' modules represent residual CNN-based filter operations. }
	\label{fig:lifting-mv}
\end{figure}

Fig.~\ref{fig:lifting-mv} illustrates the detailed structure of the prediction and update filters. For the prediction step, motion estimation between the even and odd frames $f_{2t}$ and $f_{2t+1}$ is performed to obtain the motion vectors at time instance $t$. The motion vectors are employed for motion compensation, followed by a denoising filtering module (DN). In the update step, the motion vectors are inverted to perform inverse motion compensation (MC$^{-1}$) followed by another denoising module. Due to the update step, the even frame is effectively low-pass filtered along the motion trajectory. The temporal lowpass filtering separates noise from content over time. 

Applying a denoising filter after forward and inverse motion compensation has been shown to improve compression efficiency in scalable lossless wavelet coding of dynamic CT data \cite{Lanz2019a}. This paper follows the same processing order and structure but uses trainable denoising filters, allowing for flexibility during training. The denoising filters have the same residual filter structure as the prediction and update filters of the CNN-based spatial DWT in iWave++ \cite{maiwave++}.

\subsection{Motion Estimation and Motion Vector Compression}
\noindent
The approaches for motion estimation and motion vector coding follow the state-of-the-art learned video coder DCVC-HEM~\cite{Li2022}. During motion estimation, a dense optical flow field is estimated using a Spatial Pyramid Network (SPyNet)~\cite{Ranjan2017}. With six pyramid levels, the input of SPyNet is $6\times$ downsampled. At every pyramid level, a network computes the residual flow based on the upsampled flow from the preceding level, and thus deals with relatively small motion.

To code the motion vectors obtained from SPyNet, a motion vector encoder computes a 64-channel latent representation with a 16$\times$ downscaled spatial resolution. The latents are discretized using multi-granularity quantization. The entropy model uses a hyper prior and a dual spatial prior. The latter is a two-step coding approach that exploits channel redundancies, which allows parallelization, in contrast to an autoregressive prior. The latent prior employed by DCVC-HEM conditions the entropy model on previously coded motion vector latents and is omitted for the MCTF coder. Because training is performed using only two frames, as detailed in Section~\ref{sec:training}, only one motion vector latent is available.
For more details on motion vector compression, please refer to \cite{Li2022}.
\section{Dyadic Temporal Decomposition} \label{sec:multiple}
A dyadic decomposition \cite{Taubman2002} recursively applies a wavelet transform in the temporal direction to the lowpass of the previous decomposition stage. Thus, different temporal resolutions are obtained at each decomposition level for temporal scalability. With the dyadic decomposition structure, the number of frames contained in a group of pictures (GOP) is equal to powers of two. This paper investigates GOPs containing up to eight frames.
\subsection{Coding Order and Temporal Scalability} \label{sec:codingorder}
\begin{figure}[tb]
	\centering	
	\subfloat{%
		\includegraphics[width=0.49\textwidth]{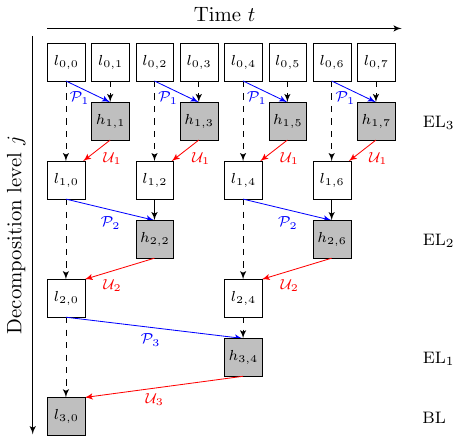}	
	}
	\caption{Coding order for a GOP size of 8. The temporal lowpass and highpass subbands are denoted as $l_{j, t}$ and $h_{j, t}$. The gray frames are coded from temporal decomposition level 1 to 3. }
	\label{fig:gop8}
\end{figure}

\noindent 
Fig.~\ref{fig:gop8} illustrates the coding order of MCTF for a GOP consisting of 8 frames.  
Because MCTF is an open-loop structure, motion estimation is performed on the original frames instead of the decoded frames. In the first temporal decomposition level, the operator $\mathcal{P}_1$ predicts all odd-indexed frames from the respective preceding frame. The resulting four temporal highpass frames $h_{1, t}$ and their corresponding motion vectors can be directly coded.  Next, the temporal lowpass frames are obtained from the update operation $\mathcal{U}_1$ which receives the highpass frames as input. After the first temporal decomposition level, there are four temporal lowpass frames. MCTF repeats this decomposition in the temporal direction until only the single temporal lowpass frame $l_{3, 0}$ in decomposition level $j=3$ is left. Overall, the highpass frames $h_{1, t}$ from the first decomposition level can be coded first, followed by the highpass frames from the deeper temporal levels. Finally, the lowpass frames $l_{3, 0}$ from the lowest temporal decomposition level are transmitted.

Note that the distance between frames $d$ in the temporal direction increases with every temporal decomposition level $j$ according to $d=2^{j-1}$. Hence, the frame distance $d$ is equal to $4$ in temporal decomposition level $j=3$. This is disadvantageous in terms of rate-distortion performance compared with regular P frame coding with a frame distance of $d=1$ for every P frame. However, MCTF has the benefit of providing temporal scalability: the lowpass subbands are similar to the original sequence and therefore correspond to a Base Layer (BL). The highpass subbands contain residual information that serves as an Enhancement Layer (EL). The further the input video sequence is decomposed in the temporal direction, the more ELs are available. For a GOP size of 8, there are three ELs as indicated in Fig.~\ref{fig:gop8}.
Owing to the different temporal decomposition levels,  dedicated MCTF filtering, motion estimation, and motion vector compression networks for each temporal decomposition level are beneficial. The benefits of the different MCTF stages are evaluated in Section~\ref{sec:ablation}.

On the decoder side, the inverse MCTF is performed by reversing the order of the prediction and update filters. 

\subsection{Training Strategy and Loss}  \label{sec:training}
\noindent
This paper adopts a multi-stage training strategy, of which Table~\ref{tab:training} provides an overview. During the entire training procedure, each training sample consists of two frames. In the first part, a single MCTF stage is trained (training stage 1-3), and more stages are added in the second training part (training stage 4-5) depending on the GOP size. Thus, dedicated models for different GOP sizes are trained to consider the varying number of temporal decomposition levels. The two iWave++ models employed for coding the temporal lowpass and highpass subbands are initialized with models pretrained on image data.

\subsubsection{First training part: Single MCTF stage}
During the first two training stages, only the network components for MCTF are trainable. They consist of motion estimation, motion vector compression, and DN modules. In the first stage, the loss is the distortion  $D_{\mathrm{ME}}$ between the frame to be predicted $f_{2t+1}$ and the prediction $\mathcal{P}_1(f_{2t})$. The second stage additionally considers the rate  $R_{\mathrm{MV}}$ required for motion vector coding. In the next stage, the entire network is trainable and the loss is the regular rate-distortion loss.
\bgroup
\def\arraystretch{1.4}

\begin{table}[tb]
	\caption{Training schedule for a GOP size of 4/8. A training sample consists of two frames in each training stage. LR denotes the learning rate, $d_{\mathrm{max}}$ the maximum frame distance between two frames in a training sample, and ''parts'' refers to the trainable components of the network. ''All'' parts include the MCTF stages and the iWave++ models.   }
		\renewcommand{\arraystretch}{1.5}
	\begin{center}
	\resizebox{0.5\textwidth}{!}{
		\begin{tabular}{l|l|l|l|l|l}
\multirow{2}{*}{MCTF stages}  &  \multirow{2}{*}{Parts}  & \multirow{2}{*}{$d_{\mathrm{max}}$}  & \multirow{2}{*}{Loss}  & \multirow{2}{*}{LR}  & \multirow{2}{*}{Epochs}  \\[-0.1cm]
			 &   & &  &  &  \\
			\hline
			Single & MCTF & 1 	 & $D_{\mathrm{ME}}$ & $\num{1e-04}$ & 1 \\	
			Single & MCTF & 1 	 & $D_{\mathrm{ME}}+R_{\mathrm{MV}}$  & $\num{1e-04}$ & 3 \\	
			Single & All  & 1 	 & $\mathcal{L}_{\mathrm{full}}$  & $\num{1e-05}$ & 5 \\	
			\hline
			Multiple (2/3) & MCTF  & 2/4 	 & $\mathcal{L}_{\mathrm{full}}$   & $\num{5e-05}$ & 2 \\	
			Multiple (2/3) & All  & 2/4 	 &  $\mathcal{L}_{\mathrm{full}}$   & $\num{1e-05}$ & 3 \\	
			
		\end{tabular}
		\label{tab:training}
	}
	\end{center}
\end{table}
\egroup 
The full rate-distortion loss for two frames reads:
\begin{equation}
	\mathcal{L}_{\mathrm{full}} =
	\sum_{i=0,1}	R_{\mathrm{all}, i} + \lambda \cdot  D_{\mathrm{MSE}}(f_i, \hat{f}_i), \notag
\end{equation}
where $i$ denotes the frame number and the distortion term corresponds to the Mean Squared Error (MSE) between the original frame $f_i$ and the reconstructed frame $\hat{f}_i$. $R_{\mathrm{all}, i}$ consists of the rate required to code the temporal subbands using an iWave++ model. If the corresponding frame $i$ is coded as a temporal highpass subband,  $R_{\mathrm{all}, i}$ also includes $R_{\mathrm{MV}}$. This paper considers lossy compression, where the only information loss stems from the scalar quantization operation of the iWave++ models.

\subsubsection{Second training part: Multiple MCTF stages}
To account for multiple temporal decomposition levels, multiple MCTF networks are used, where the additional MCTF stages are initialized with the parameters of the already available MCTF stage. For a GOP size of 4 with two temporal decomposition levels, two MCTF stages and a maximum frame distance $d_{\mathrm{max}}$ of two are used. For every batch element, a random frame distance between one and $d_{\mathrm{max}}$ is selected. Depending on the frame distance, a different MCTF stage with different networks is chosen. Thus, for a GOP size of 4, it is randomly alternated between optimizing the first MCTF stage with a frame distance of one and the second MCTF stage with a frame distance of two. Thereby, the different MCTF stages share the iWave++ models employed for coding the temporal lowpass and highpass subbands.

In the last two training stages, again only the MCTF components are trained first and then all network modules are jointly optimized as can be seen in Table~\ref{tab:training}. To consider inverse MCTF for multiple decomposition levels during training, experiments were conducted using four frames per batch element. However, they showed that training becomes unstable, and the final rate-distortion performance is significantly worse than training with two frames and one temporal level.

For a GOP size of 8, the number of MCTF stages is increased from two to three. The maximum frame distance $d_{\mathrm{max}}$ in the last two stages (see Table~\ref{tab:training}) is set to 4 to account for the GOP structure shown in Fig.~\ref{fig:gop8}.

\subsection{Downsampling Strategy for Temporal Displacements in MCTF}
\label{sec:ds}
\noindent

\begin{figure}[tb]
	\centering	
	\subfloat{%
		\includegraphics[width=0.49\textwidth]{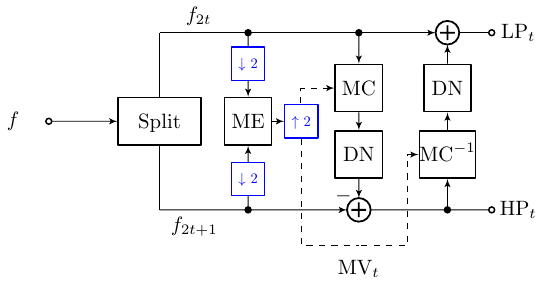}	  
	}
	\caption{Lifting structure with downsampling strategy for decomposition levels $j > 1$ (\textit{MCTF-DS}). Here, the ''ME'' module performs motion estimation on downscaled input frames. The reconstructed motion vectors MV$_t$ obtained from motion vector compression are upscaled before being used for forward and inverse motion compensation.  }
	\label{fig:lifting-mv-ds}
\end{figure}
The larger the temporal decomposition level, the larger the temporal distance between the frames in the original sequence (see Section~\ref{sec:codingorder}). Therefore, considerably larger temporal displacements are possible. If the motion is too strong for the motion estimation network to predict accurately, prediction errors can lead to ghosting and error propagation across decomposition levels.   

To address larger motion, computing and transmitting motion vectors at a lower spatial resolution for temporal decomposition levels larger than one is proposed. Specifically, the current frame and reference frame before motion estimation are downscaled by a factor of two for every temporal decomposition level $j>1$ as illustrated in Fig.~\ref{fig:lifting-mv-ds}. Hence, the motion vectors are coded at lower resolution and upscaled after the motion vector decoder. Both bilinear down- and upsampling are performed. The upscaled motion vectors are then used for the forward and inverse MCTF. The proposed downsampling strategy (\textit{MCTF-DS}) does not require additional training, and its benefits are evaluated in Section~\ref{sec:ablation}.
\subsection{Content-Adaptive MCTF (MCTF-CA)}
\label{sec:ca}
The coding efficiency of MCTF is highly dependent on the motion-compensated prediction quality as motion estimation errors propagate to higher temporal decomposition levels. Even with the downsampling strategy, the motion present in a scene can be too strong for the motion estimation network or occluded regions can limit the prediction quality. Therefore, adaptive temporal scaling for each video sequence can lead to improved coding efficiency compared with uniform dyadic temporal decomposition by mitigating ghosting and thus error propagation. Lanz~\etal \cite{Lanz2019} investigated content-adaptive wavelet lifting for scalable lossless coding of medical data by choosing the number of temporal decomposition levels based on the sequence content. This paper proposes the adoption of a content-adaptive wavelet lifting approach for our lossy wavelet video coder, which is referred to as \textit{MCTF-CA}. This approach does not require additional training.

In the following section, the concept of content-adaptive MCTF for a GOP size of 8 is explained. During inference, the coding costs for a GOP consisting of 8 frames are optimized. As a cost criterion, the rate-distortion cost for $N=8$ frames is evaluated as:
 \begin{equation}
 	\mathcal{C}_{N}=
 	\sum_{i=0}^{N-1}	R_{\mathrm{all}, i} + \lambda \cdot  D_{\mathrm{MSE}}(f_i, \hat{f}_i), \notag
 \end{equation}
where the tradeoff parameter $\lambda$ is chosen according to the value employed for training the MCTF model. With one MCTF model trained for a GOP size of 8, evaluate different options for coding the current GOP. Subsequently, the variant with the minimum coding cost is chosen:
\begin{equation}
	\min \left( \mathcal{C}_{8, \mathrm{GOP}8}, \, \mathcal{C}_{8, \mathrm{GOP}8}^{\mathrm{DS}}, \, \mathcal{C}_{8, \mathrm{GOP}4}, \, \mathcal{C}_{8, \mathrm{GOP}4}^{\mathrm{DS}}, \, \mathcal{C}_{8, \mathrm{GOP}2}  \right), \notag
\end{equation}
where the notation $\mathcal{C}_{8, \mathrm{GOP}4}^{\mathrm{DS}}$ denotes the cost of coding 8 frames in smaller GOPs of size 4 with the downsampling strategy, for example. Either a GOP size of 8 is coded or split into several smaller GOPs. Here, two GOPs of size 4 or four GOPs of size 2 are possible. In addition, it is decided whether to use the downsampling strategy or not.

In total, five options are considered for coding a GOP with 8 frames. The choice for each GOP needs to be transmitted to the decoder side. However, the overhead of transmitting three bits per eight frames is negligible. Hence, binary encoding is used to signal the content-adaptive choice for a coding unit with eight frames.

For a GOP size of 4, there are three options: Two GOPs of size 2 and one GOP of size 4, with or without downsampling.

\section{Experiments and Results} 
\label{sec:experiments}
\subsection{Experimental Setup}
\noindent \vspace{-0.7cm}
\subsubsection{Training details} 
The networks described above are implemented using the PyTorch framework.
The Vimeo90K data set \cite{xue2019video} is used for training and the batch size is set to 8. During training,  patches of size $128 \times 128$ are cropped from the luma channel of the respective training sample, whereas no cropping is performed during inference. By choosing the rate-distortion trade-off parameter according to $\lambda =  \left\{ 0.007, 0.01, 0.03, 0.05, 0.08 \right\}$, five models are obtained for each GOP size. AdamW \cite{loshchilov2018decoupled} is used as optimizer. Furthermore, the iWave++ models pretrained on luma data from \cite{icassp2023} are used for temporal subband coding. SPyNet \cite{Ranjan2017} is initialized with the ''sintel-final'' model\footnote{https://github.com/sniklaus/pytorch-spynet} trained on a synthetic data set. 

As described in Section~\ref{sec:training}, separate models with multiple MCTF stages are trained for GOP sizes of 4 and 8, because the seven frames available in sequences from the Vimeo90K data set allow considering up to three temporal decomposition levels, that is, a maximum GOP size of 8. In line with the MCTF evaluation setup from Dong~\etal \cite{Dong2022} with a GOP size of 8, it is shown that in this setting, MCTF performs competitive to state-of-the-art coders. 
\subsubsection{Test conditions}
The UVG \cite{Mercat2020} and MCL-JCV \cite{Wang2016} data sets are used for testing. The sequences in both data sets have a resolution of 1920$\times$1080 and are in YUV 4:2:0 format. UVG consists of 7 sequences and MCL-JCV of 30. To consider a different resolution of 1280 $\times$ 720, the \mbox{JCT-VC} class E data set (HEVC E) containing three YUV 4:2:0 sequences are used. The test conditions in \cite{Li2022} are followed by evaluating on the first 96 frames of each sequence. In addition, the evaluation includes three sequences from the UVG 4K \cite{Mercat2020} data set (\textit{CityAlley}, \textit{FlowerFocus}, \textit{FlowerKids}) with a resolution of 3840$\times$2160 and testing is performed on the first 24 frames.

DCVC-HEM\footnote{https://github.com/microsoft/DCVC/tree/main/DCVC-HEM} \cite{Li2022} and DCVC\footnote{https://github.com/microsoft/DCVC/tree/main/DCVC} \cite{li2021deep} are evaluated with GOP sizes of 4 and 8 for a fair comparison with the MCTF approach. Thereby, publicly available models from the authors are used, which were trained on Vimeo90K. As a traditional hybrid video coder, HM 16.25\footnote{https://vcgit.hhi.fraunhofer.de/jvet/HM/-/releases/HM-16.25} is included. HM is used in the Lowdelay P (LD-P) configuration because the learned video coders only support unidirectional motion estimation. HM is evaluated in its default main profile with an intra period and GOP sizes of 4 and 8 as well.

The evaluation is performed in terms of RGB-PSNR, as this is common in learned video compression, and the aim of this paper is to provide comparable measurements. 
The MCTF approach and HM receive the input video sequence in YUV 4:2:0 format, whereas the input is converted to RGB 4:4:4, as required by DCVC-HEM and its predecessor DCVC. The wavelet video coder supports input data in YUV 4:2:0 format as well as in 4:4:4 format, because the color channels are coded independently by iWave++. The motion vectors are computed based on the luma channel. They are re-used for the chroma channels, and bilinear downsampling is performed if necessary.

\subsection{Experimental Results}
\noindent \vspace{-0.9cm}
\subsubsection{Comparison to state-of-the-art video coders}
\begin{figure}[tb]
	\includegraphics[width=0.43\textwidth]{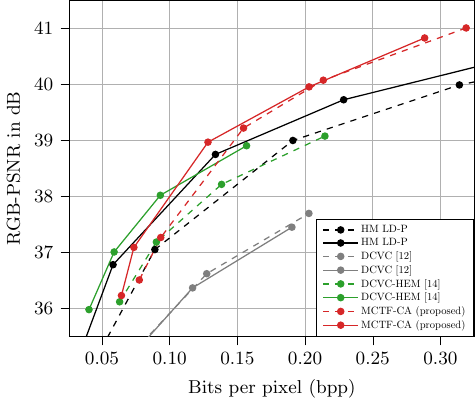}
	\caption{Rate-distortion evaluation on the UVG data set. Solid lines correspond to a GOP size of 8 and dashed lines to a GOP size of 4.}
	\label{fig:uvg2}
\end{figure}

\begin{figure}[tb]
	\includegraphics[width=0.43\textwidth]{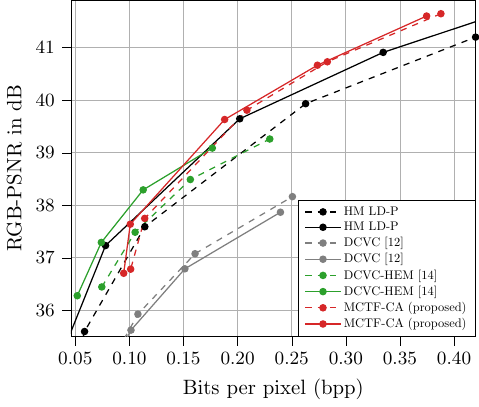}
	\caption{Rate-distortion evaluation on the MCL-JCV data set. Solid lines correspond to a GOP size of 8 and dashed lines to a GOP size of 4.}
	\label{fig:mcl-jcv}
\end{figure}

\begin{figure}[tb]
	\includegraphics[width=0.43\textwidth]{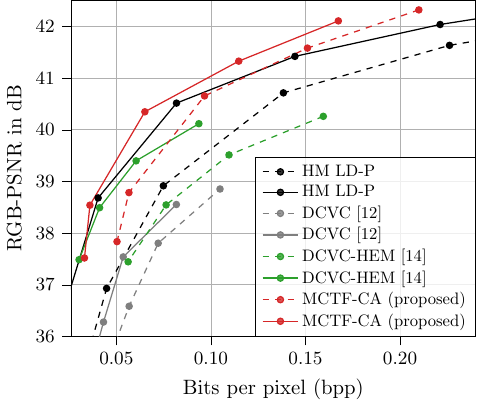}
	\caption{Rate-distortion evaluation on the HEVC E data set. Solid lines correspond to a GOP size of 8 and dashed lines to a GOP size of 4.}
	\label{fig:hevc}
\end{figure}

\begin{figure}[tb]
	\includegraphics[width=0.42\textwidth]{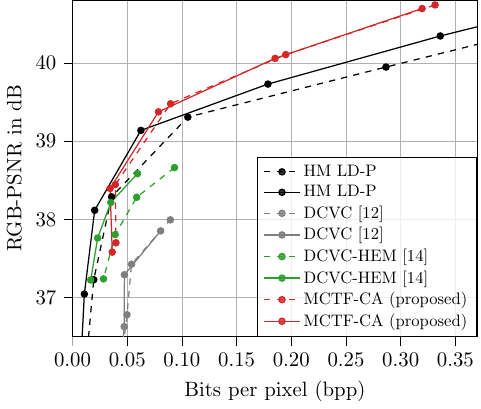}
	\caption{Rate-distortion evaluation on 3 sequences (\textit{CityAlley,} \textit{FlowerFocus, FlowerKids}) from the UVG 4K data set. Solid lines correspond to a GOP size of 8 and dashed lines to a GOP size of 4.}
	\label{fig:uvg_4k}
\end{figure}

\paragraph*{Rate-distortion curves} \mbox{} \\
\noindent The novel approach is compared to HM, the state-of-the-art learned video coder DCVC-HEM~\cite{Li2022}, and its predecessor DCVC~\cite{li2021deep}. Figs.~\ref{fig:uvg2}-\ref{fig:uvg_4k} show the rate-distortion curves for the UVG, MCL-JCV, HEVC E, and UVG 4K data sets, respectively. The dashed lines correspond to a GOP size of 4, whereas solid lines indicate a GOP size of 8. 
 \noindent

Clearly, the conditional coder DCVC (gray) is not competitive with the remaining video coders. The approach performs better for a smaller GOP size of 4 compared to a GOP size of 8 on two data sets, which implies an error propagation issue. Its successor, DCVC-HEM (green), on the other hand, can effectively exploit temporal redundancies for the larger GOP size of 8. DCVC-HEM outperforms HM in lower bitrate ranges on UVG and MCL-JCV, whereas HM always performs better at higher rates. 

The rate-distortion performance of the best-performing model, MCTF-CA (red), behaves in the opposite way: the higher the rate, the better the approach performs relative to HM. At higher rates, the model clearly outperforms HM for all data sets and GOP sizes. The performance degrades only for the MCTF-CA model at the lowest rate point ($\lambda = 0.007$) compared with the other rate points. The MCTF-CA model performs particularly well at high rates, owing to its invertible wavelet transforms. The perfect reconstruction property allows lossless compression without quantization, and therefore provides the capacity for high coding efficiency at high quality.

\paragraph*{Bj{\o}ntegaard Delta rate} \mbox{}  \\ 
\noindent For a quantitative evaluation of the rate-distortion performance, the Bj{\o}ntegaard Delta (BD) rate savings of the learned video coders are measured using HM LD-P as an anchor. Note that the BD values need to be handled with caution because the available supporting points of DCVC-HEM and DCVC cover a limited bitrate and quality range. Thus, comparisons in terms of the BD metric can be less reliable \cite{Herglotz2023}, and rate-distortion curves should be considered to obtain a complete picture. Therefore, using HM as an anchor avoids comparing the two conditional coders with the proposed method directly, but still perform comparisons over different bitrate and quality ranges. To cover the entire bitrate-distortion range of the learned video coders, HM is evaluated with Quantization Parameters (QP) values $\mathrm{QP} = \left\{ 32, 27, 22, 19, 17, 15, 13\right\}$. The integration area for BD rate calculation is determined by the respective learned video coder, that is, by the minimum and maximum RGB-PSNR values obtained with the learned coder. Compared with the entire rate-distortion curve of HM, the overlap of the rate-distortion curve of DCVC-HEM with respect to the bitrate lies in the range of 14-29\%. The overlap in terms of RGB-PSNR is between 42 and 46\% depending on the data set and GOP size. Comparing the overlap of the rate-distortion curves of HM and MCTF-CA, the rate overlap is between 36 and 66\%, whereas the distortion overlap ranges from 70 to 94 \%. Hence, the MCTF models cover a larger rate-distortion range, as shown in Figs.~\ref{fig:uvg2}-\ref{fig:uvg_4k}.
\begin{table}[tb]
	\centering
	\caption{Rate-distortion evaluation on the UVG, MCL-JCV, HEVC E, and UVG 4K data sets for different GOP sizes. Average BD rate savings are provided relative to HM in LD-P configuration as an anchor.}
	\renewcommand{\arraystretch}{1.3}

	\centerline{(a) \textbf{GOP 4} \vspace{0.2cm}}
	\begin{tabular}{m{1.7cm} |m{1.7cm}|m{1.7cm}|m{1.7cm}}
				    & \small DCVC   				& \small DCVC-HEM				& \small MCTF-CA  \\	\hline
		UVG 	    & 	 +64.10\%  	& 	-0.75\% 		&    \textbf{-21.48\%} \\
		MCL-JCV   	& 	+68.27\% 	& 	 -1.42\% 		&  \textbf{-12.63\%}   \\
		HEVC E   	& +35.83\% 		& 	+17.84\%		&  \textbf{-26.20\%} \\ 
		UVG 4K 		& 	+207.75\% 	& +56.53\%			&  \textbf{-21.48\%}   \\
	\end{tabular}\\[0.2cm]

	\centerline{(b) \textbf{GOP 8} \vspace{0.2cm}}
	\begin{tabular}{m{1.7cm} |m{1.7cm}|m{1.7cm}|m{1.7cm}}
				& \small DCVC   				& \small DCVC-HEM				& \small MCTF-CA  \\	\hline
	UVG 	    & +134.50\%				&  -3.93\%				&  \textbf{ -9.17\%}  \\
	MCL-JCV   	&  +131.30\% 			&  \textbf{-4.36} \%	&  -0.41\%  \\
	HEVC E   	& +96.68\%				&  +13.31\%				& \textbf{-11.22\%}  \\ 
	UVG 4K 		& +405.48\%				& 	+52.31\% 			&  \textbf{ +3.70\%  }\\
	
\end{tabular}
	\label{tab:bd}
	\vspace{0.5cm}
\end{table}

Table~\ref{tab:bd} contains the BD measurements for all four data sets and for both GOP sizes. Over the entire bitrate range, \mbox{DCVC-HEM} performs best on the MCL-JCV data set for a GOP size of 8, achieving a BD rate reduction of approximately \mbox{-4\%} compared to HM. In the remaining cases, MCTF-CA performs the best. It achieves BD rate savings of up to -21\% and -9\% on the UVG data set for GOP sizes of 4 and 8, respectively. On MCL-JCV, BD rate savings of -12\% are obtained for a GOP size of 4. Furthermore, MCTF-CA achieves coding gains of -26\% and -11\% for GOP sizes of 4 and 8, respectively, on HEVC E. Overall, the high-resolution sequences from the UVG 4K data set are the most challenging for all learned video coders. MCTF-CA only achieves coding gains over HM for a GOP size of 4, but nevertheless performs favorably in comparison to the remaining learned coders.

\paragraph*{Per-sequence evaluation on the UVG data set} \mbox{} \\ \noindent
The coding performance of DCVC-HEM and MCTF-CA is assessed for each of the seven sequences in the UVG data set. Table~\ref{tab:perSeq} provides BD rate savings relative to HM as an anchor. Independent of the GOP size, DCVC-HEM performs better than the proposed approach compared to HM for sequences with stronger motion, namely, \textit{Jockey}, \textit{ReadySteady}, and \textit{YachtRide}. These sequences mostly contain relatively large translational motion. In contrast, the MCTF-CA approach performs best for sequences with high spatial detail and more irregular motion. For example, the approach achieves BD rate savings of over -63\% for the \textit{Beauty} scene, which is challenging because of moving hair. Here, \mbox{DCVC-HEM} struggles and is the least efficient compared to HM. 

\begin{figure}[tb]
	\centering	
	\subfloat[$l_{0,0}$ (original frame 0)]{%
		\includegraphics[width=0.27\textwidth]{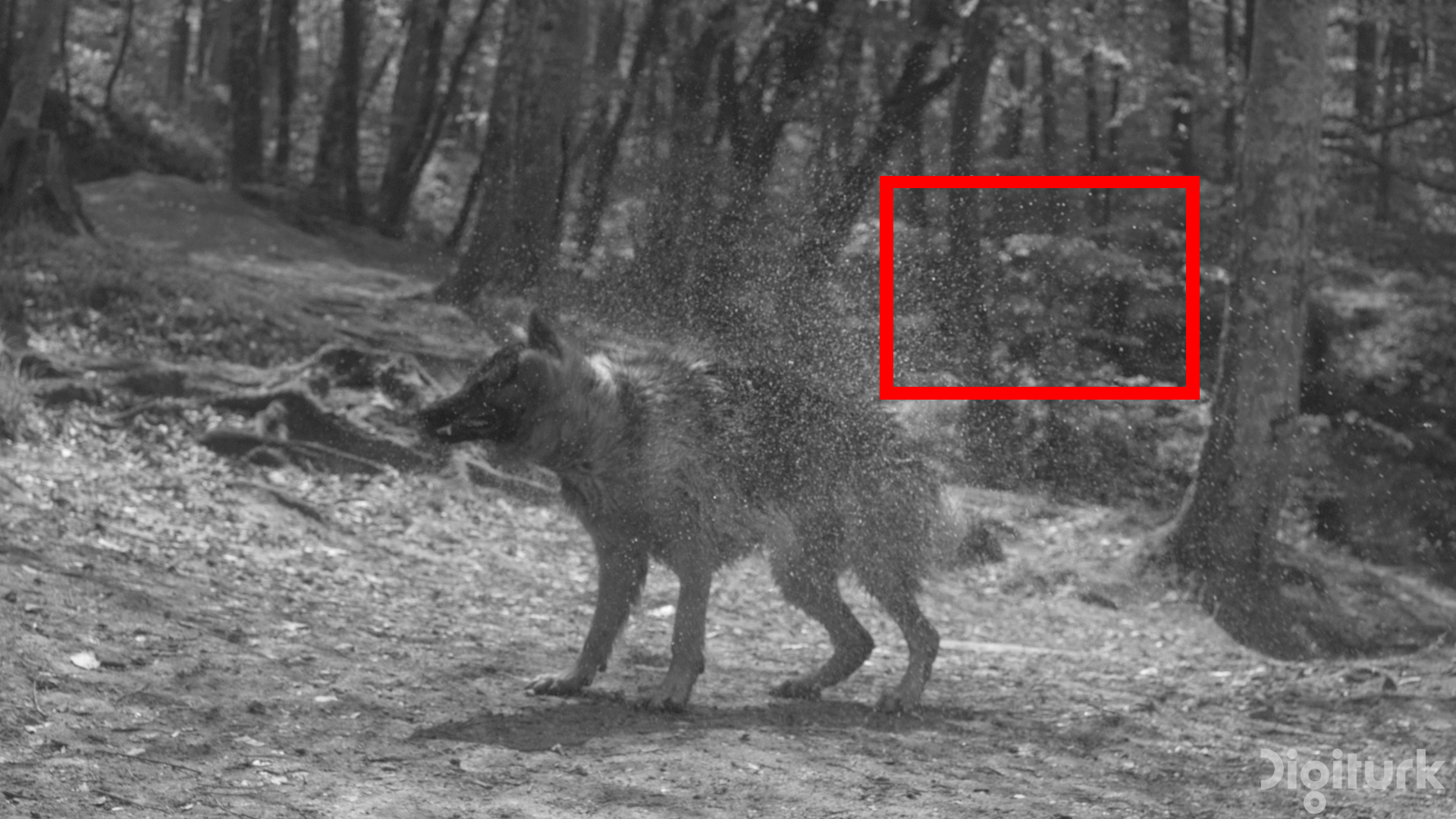}	  
	}
	\subfloat[$l_{0,0}$ (crop)]{%
		\includegraphics[width=0.2\textwidth]{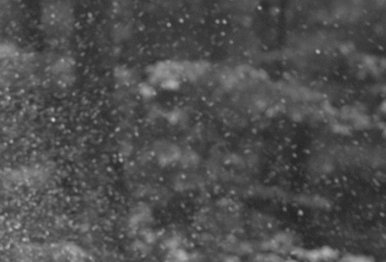}	  
	}\\
	\subfloat[$l_{2,0}$ ($\mathcal{U}_1$ \& $\mathcal{U}_2$ applied)]{%
		\includegraphics[width=0.2\textwidth]{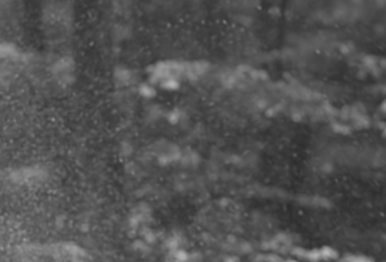}	  
	}
	\subfloat[$h_{3,4}$ (= $\mathcal{P}_3 (l_{2,0})$)]{%
		\includegraphics[width=0.2\textwidth]{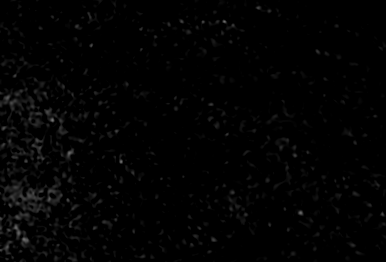}	  
	}\\
	\subfloat[$h_{2,2}$ ]{%
	\includegraphics[width=0.2\textwidth]{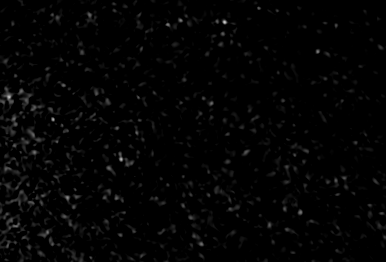}	  
}
	\subfloat[$h_{1,1}$ ]{%
		\includegraphics[width=0.2\textwidth]{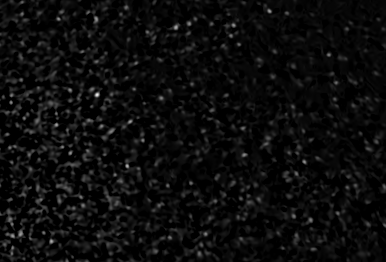}	  
	}
	\caption{Impact of the temporal update operation. Subfig.~(a) shows the first frame of the \textit{ShakeNDry} sequence from the UVG data set.  (d)-(f) depict temporal highpass frames coded in different temporal decomposition levels by a MCTF-CA model ($\lambda=0.08$, GOP size 8). The highpass frame $h_{3,4}$ from the highest temporal decomposition level has less prediction errors compared to $h_{2, 2}$ and $h_{1, 1}$ (black corresponds to zeros). This is because $h_{3,4}$ is predicted from $l_{2, 0}$ shown in (c). Here, the application of temporal updates to $l_{2, 0}$ improves the prediction and thus coding efficiency. }
	\label{fig:update}
\end{figure}

Overall, the per-sequence evaluation shows that MCTF leads to superior coding performance compared to an "IPPP$\hdots$" coding order for specific scene contents. The following example of the \textit{ShakeNDry} sequence illustrates the benefits of the temporal update operation. The scene has a static background, but contains challenging motion with flying water drops. With the MCTF-CA model, the first GOP of the sequence is coded with a GOP size of 8, that is, three temporal decomposition levels. The temporal updates help improve the coding efficiency of the temporal highpass frames at higher temporal decomposition levels: the highpass frames in the first, second, and third level require 0.38~bpp, 0.27~bpp, and 0.20~bpp at approximately 42.2~dB.  As shown in Fig.~\ref{fig:update}(d)-(f), the highpass $h_{3, 4}$ from temporal decomposition level three contains fewer prediction errors compared to the other levels, which leads to better coding efficiency. The application of two temporal update operations (see Fig.~\ref{fig:update} (c)) creates a better representation for the prediction compared to the original frame in Fig.~\ref{fig:update} (b) through lowpass filtering along the motion trajectory.
\begin{table}[tb]

	\caption{BD rate savings for each of the 7 UVG sequences over HM in LD-P configuration.}
	\renewcommand{\arraystretch}{1.2}
	\resizebox{0.49\textwidth}{!}{ 
	\begin{tabular}{l|c|c|| c| c}

			&	\multicolumn{2}{c||}{{GOP 4}} & \multicolumn{2}{c}{{GOP 8}} \\
											&	 DCVC-HEM  & MCTF-CA &  DCVC-HEM  & MCTF-CA\\
		\hline
		HoneyBee							&	+16.72\% 	&   -52.38\%& +17.60\% &  -50.29\%\\
		Bosphorus							&  -9.27\%		& -14.71\%&  -16.27\% &   -9.87\%\\
		Beauty 								&	+337.29\%	&  -64.37\%&  +373.47\% &   -63.63\%\\
		YachtRide							&-18.21\%	& -5.69\%&-22.40\% &  +1.48\%\\
		ShakeNDry							&-3.52\%		& -29.06\%& -0.46\% &  -21.44\%\\
		Jockey 								&+12.78\%	&  +18.81\%	& +10.16\% &   +93.19\%\\
		ReadySteady							& -12.17\%  &  -10.93\%&  -19.35\% &  +14.16\%\\
	\end{tabular}

	}
	\label{tab:perSeq}
\end{table}

\begin{figure}[tb]
	\centering	
	\includegraphics[width=0.43\textwidth]{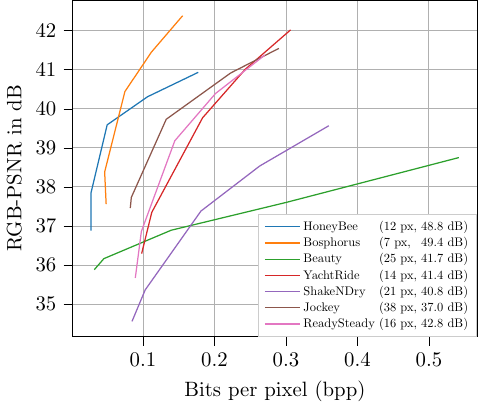}	  

	\caption{Comparison of the rate-distortion curves of MCTF-CA (GOP size of 8) for every sequence of the UVG data set. For each sequence, the motion strength in pixels (px) and motion-compensated prediction quality in dB averaged over all frames are provided. For these measurements, the motion vectors between successive frames required for motion compensation are estimated using a SPyNet model. Thereby, the motion strength for a single frame is measured as the maximum motion vector length in horizontal or vertical direction.  }
	\label{fig:perSeq}
\end{figure}

When comparing the rate-distortion curves of MCTF-CA for every sequence of the UVG data set (cf. Fig.~\ref{fig:perSeq}), the \textit{ShakeNDry} sequence is one of the most challenging sequences next to the \textit{Beauty} sequence. Fig.~\ref{fig:perSeq} provides the motion-compensated prediction quality in terms of PSNR and maximum motion vector length in pixels averaged over all 96 evaluated frames for each sequence. These values are computed using a SPyNet model trained on Vimeo90K without considering motion vector compression. These measurements show that a high prediction quality of over 48 dB and relatively small motion (\textit{HoneyBee}, \textit{Bosphorus}) are associated with the best rate-distortion performance of MCTF-CA. However, a lower prediction quality and larger motion do not necessarily lead to poor rate-distortion performance; for example, MCTF-CA performs better on the \textit{Jockey} sequence than on the \textit{Beauty} sequence because factors such as high spatial detail contained in a sequence influence the coding efficiency as well.

\vspace{-0.4cm}
\paragraph*{Complexity} \mbox{} \\ \noindent
The computational complexity of the MCTF-based approach is assessed in terms of model size and kilo multiply-accumulate operations per pixel (kMAC/px). As shown in Table~\ref{tab:complexity}, the MCTF-CA  approach is more complex with respect to both model size and kMACs/px. Note that most of the model complexity of MCTF-CA is attributed to the temporal subband coder iWave++. For a GOP size of 8, the MCTF modules only account for 29~\% of the model size and 12~\% of the required kMACs/px. Because of the dedicated MCTF stages for every temporal decomposition level, the MCTF modules have a larger influence on the model size relative to MACs.
\begin{table}[tb]
	\caption{Complexity comparison of learned video coders for an input size of 1920$\times$1080 in terms of model size and kilo multiply-accumulate operations per pixel (kMAC/px).}
	\renewcommand{\arraystretch}{1.2}
		\begin{tabular}{p{1.7cm}|p{1.7cm}| p{1.7cm}|p{1.7cm}}

  & \small DCVC & \small DCVC-HEM & \small MCTF-CA \\
  \hline

Model size		& 32 MB &   70 MB& 90 MB\\
kMAC/px	& 1167 & 1673 & 3554\\
		\end{tabular}
	\label{tab:complexity}
\end{table}


\subsubsection{Ablation Study: MCTF configuration}
\label{sec:ablation}
In the following section, several MCTF coder configurations are examined. In doing so, the benefits of the proposed downsampling strategy and content-adaptive MCTF approach are evaluated.
\begin{table}[tb]
	\caption{Rate-distortion evaluation on the UVG and MCL-JCV data sets for different GOP sizes. Average BD rate savings are provided relative to the baseline MCTF model as an  anchor.}
	\renewcommand{\arraystretch}{1.5}
		\resizebox{0.5\textwidth}{!}{
		\begin{tabular}{l|c|c||c|c}
			&	\multicolumn{2}{c||}{{GOP 4}} & \multicolumn{2}{c}{{GOP 8}} \\
			\hline
				  & UVG   				& MCL-JCV   			& UVG   			 & MCL-JCV  \\	\hline
	MCTF-Single	  & +17.26\%  			&  +16.65\% 			& +34.51\% 		 &  +29.60\% \\
	MCTF-DS 	  & -2.90\%  			&  -2.11\% 			& -3.46\% 		 	 &  +1.05 \% \\
	MCTF-CA 	  &  \textbf{-4.68\%}  &   \textbf{-8.74\%} 	& \textbf{ -10.15\%} &  \textbf{-14.94\% } \\
		\end{tabular}
	\label{tab:bd2}
}
\end{table}

\begin{table}[tb]
	\caption{BD rate savings for each of the 7 UVG sequences over the baseline MCTF model as an anchor.}
	\renewcommand{\arraystretch}{1.2}
	\vspace{0.2cm}
	\resizebox{0.5\textwidth}{!}{ 
		\begin{tabular}{l|c|c|| c| c}
			
			&	\multicolumn{2}{c||}{{GOP 4}} & \multicolumn{2}{c}{{GOP 8}} \\
							&	 MCTF-DS  & MCTF-CA &  MCTF-DS  & MCTF-CA\\
			\hline
HoneyBee					&	+0.25\% 	& -0.33\% 	& +9.83\% 		&  -0.01\%\\
Bosphorus					&   -1.76\%	& -2.36\%	& -0.02\% 		&   -0.02\%\\
Beauty 						&	-0.93\%	& -1.99\%	&  +3.45\% 	&   -8.31\%\\
YachtRide					&   -4.22\%	& -4.22\%	& -5.01\% 		&  -12.17\%\\
ShakeNDry					&   -0.81\%	& -0.81\%	& +0.34\% 		&  -0.13\%\\
Jockey 						&   -7.17\%	& -16.44\%	& -7.36\% 		&   -25.15\%\\
ReadySteady					&   -3.86\%  	& -4.39\%	& -12.39\% 	&  -12.64\%\\
		\end{tabular}
	}
	\label{tab:perSeq2}
\end{table}

\vspace{-0.3cm}
\paragraph*{Multiple MCTF Stages}\mbox{} \\ \noindent
First, a single MCTF stage (''MCTF-Single'') is evaluated and compared with multiple MCTF stages. The latter uses dedicated MCTF modules for each temporal decomposition level, that is, different DN, motion estimation, and motion vector compression networks for every level. Table~\ref{tab:bd2} compares the MCTF-Single model with the MCTF model with multiple MCTF stages as an anchor. The MCTF-Single model is obtained at the end of training stage three (cf. Table~\ref{tab:training}). It is included in the evaluation, because it corresponds to the standard approach commonly used in traditional MCTF.

On both data sets, MCTF-Single results in a BD rate degradation of over +16\% and +29\% for GOP sizes of 4 and 8, respectively. Therefore, multiple MCTF stages are necessary to achieve improved rate-distortion performance for higher temporal decomposition levels with larger frame distances.

\begin{figure}[tb]
	\includegraphics[width=0.49\textwidth]{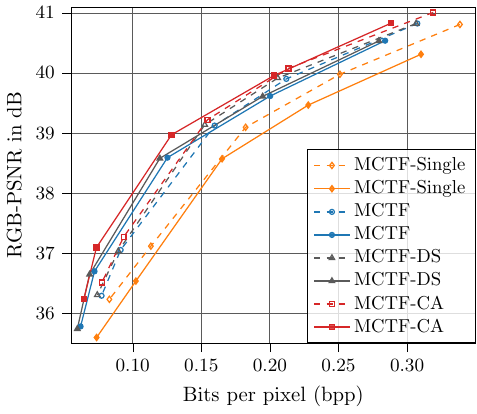}
	\caption{Rate-distortion evaluation on the UVG data set. Solid lines correspond to a GOP size of 8 and dashed lines to a GOP size of 4. \textit{MCTF-Single}: Same MCTF stage for all temporal decomposition levels. \textit{MCTF}: Different MCTF stages for each level. \textit{MCTF-DS}: Different MCTF stages with downsampling strategy during inference. \textit{MCTF-CA}: Content adaptive MCTF. \textit{Best to be viewed enlarged on a screen.}}
	\label{fig:uvg1}
\end{figure}

The impact of multiple MCTF stages on the rate-distortion curves for the UVG data set is illustrated in Fig.~\ref{fig:uvg1}. The models with multiple MCTF stages (blue) clearly outperform a single stage (orange), independent of the GOP size. 
\paragraph*{Downsampling Strategy (MCTF-DS)} \mbox{} \\ \noindent
Next, the MCTF-DS approach introduced in Section~\ref{sec:ds} is evaluated. On average, the MCTF-DS models (gray) lead to a reduced bitrate at approximately the same quality as the baseline models (blue), as shown in Fig.~\ref{fig:uvg1}. The bitrate savings are due to the smaller spatial resolution of the motion vectors, which requires a lower rate. At the same time, there is no significant quality degradation, and for some rate points, the quality is even slightly improved. 
On average, MCTF-DS leads to coding gains between 2 and 3\%, measured in terms of BD rate, compared to the MCTF model with multiple stages as an anchor (cf.  Table~\ref{tab:bd2}). However, MCTF-DS degrades the performance on the MCL-JCV data set for a GOP size of 8.

Table~\ref{tab:perSeq2} provides the BD rate evaluation for each sequence of the UVG data set. For scenes containing larger motion (\textit{ReadySteady}, \textit{YachtRide}, \textit{Jockey}), MCTF-DS achieves BD rate savings of up to -4\% and -12\% for a GOP size of 4 and 8, respectively, compared to the MCTF model with multiple MCTF stages. Consequently, MCTF-DS improves the performance for larger motion. For the \textit{HoneyBee} sequence with a small moving object and high spatial detail, the downsampling strategy leads to BD rate increases of 0.5\% and 10\% for GOP sizes of 4 and 8, respectively. This shows that although the downsampling strategy leads to improved performance for most sequences, a content-adaptive mechanism is required. 

\paragraph*{Content-Adaptive MCTF (MCTF-CA)} \mbox{} \\ \noindent
\begin{figure}[tb]
	\centering	
	\subfloat[$l_{0,0}$  (original frame 0)]{%
		\includegraphics[width=0.35\textwidth]{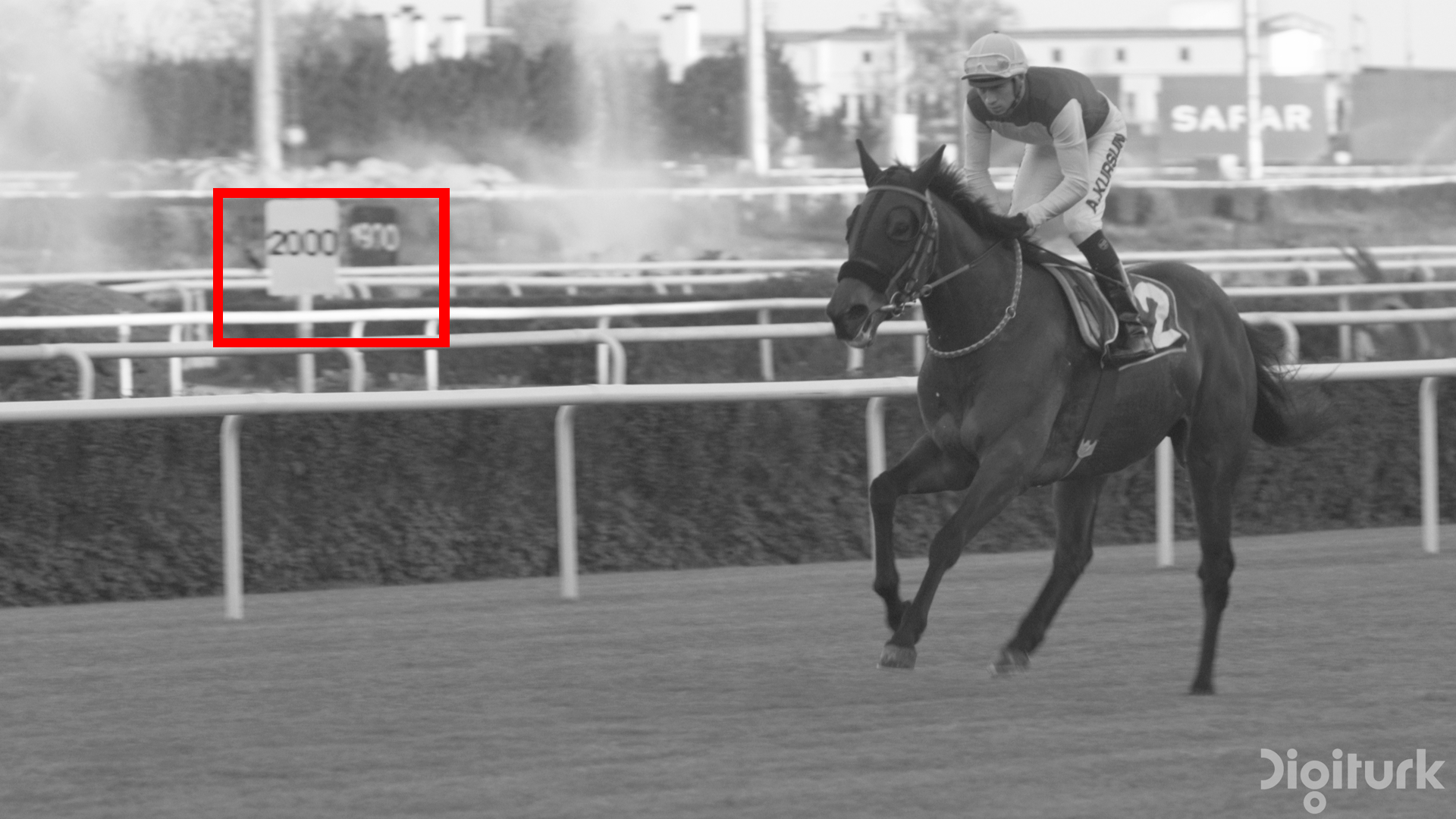}	  
	}\\
	\subfloat[$l_{0,0}$  (crop)]{%
		\includegraphics[width=0.24\textwidth]{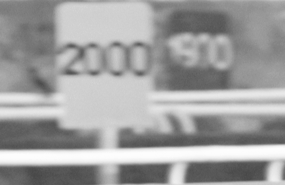}	  
	}\\
	\subfloat[MCTF: $l_{3,0}$ ]{
		\includegraphics[width=0.24\textwidth]{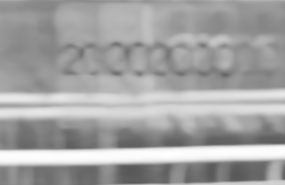}	  
	}
	\subfloat[MCTF-CA: $l_{1,0}$]{
		\includegraphics[width=0.24\textwidth]{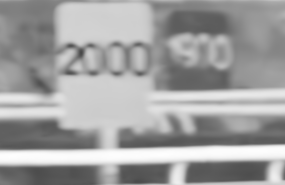}	  
	}
	\caption{Content adaptive MCTF prevents ghosting. Subfig. (a) shows the first frame of a GOP of size 8 from the \textit{Jockey} sequence. The MCTF model in (c) codes the first frame as $l_{3,0}$ in third temporal decomposition level, which leads to ghosting due to the large motion in the scene ($\mathcal{C}_{8, \mathrm{GOP}8}=1.74$).  MCTF-CA in (d) mitigates ghosting by choosing a GOP size of 2 and transmitting the first frame as $l_{1,0}$ in the first temporal decomposition level ($\mathcal{C}_{8, \mathrm{GOP}2}=1.57$). }
	\label{fig:ghosting}
\end{figure}
The MTCF-CA approach explained in Section~\ref{sec:ca} overcomes the disadvantages of MCTF-DS for some motion types and scene contents. As can be seen in Table~\ref{tab:bd2}, MCTF-CA performs best on all data sets and GOP sizes. In particular,  for a GOP size of 8, MCTF-CA provides average BD rate savings of at least 10\% compared to the MCTF model with multiple MCTF stages as an anchor. 

A detailed evaluation on every sequence of the UVG data set provided in Table~\ref{tab:perSeq2} shows that for a GOP size of 4, \mbox{MCTF-CA} improves over MCTF-DS for 5 out of 7 sequences. For the remaining two sequences, MCTF-DS is already optimal. However, for a GOP size of 8 more options for MCTF-CA are available and MCTF-DS is only optimal for the \textit{Bosphorus} sequence, which contains relatively easy translational motion. For the remaining sequences, a content-adaptive approach leads to considerable improvements in terms of BD rate; for example, MCTF-CA achieves BD rate savings of -12\% and -25\% on the \textit{YachtRide} and \textit{Jockey} sequences, respectively. Furthermore, MCTF-CA prevents the use of the downsampling strategy for sequences where it degrades rate-distortion performance, for example, for the \textit{HoneyBee} and \textit{Beauty} sequences containing high spatial detail. Therefore, content-adaptive temporal scaling is clearly advantageous in terms of rate-distortion performance, because the motion types are highly dependent on the scene content.

Fig.~\ref{fig:ghosting} provides an example of the benefit of MCTF-CA: the \textit{Jockey} sequence from the UVG data set contains strong motion, which leads to ghosting for some GOPs (cf. Fig.~\ref{fig:ghosting}(c)) when processing the sequence with a uniform temporal decomposition, that is, a constant GOP size of 8 with the MCTF model. MCTF-CA adaptively chooses a smaller GOP size if ghosting harms the coding costs. As can be seen in Fig.~\ref{fig:ghosting}(d), MCTF-CA prevents ghosting by determining a GOP size of 2, which can be coded most efficiently. 
\section{Conclusion}
\label{sec:conclusion}
This paper introduced the first end-to-end trainable wavelet video coder based on MCTF. It presented a training strategy that considers multiple temporal decomposition levels during training. Moreover, a downsampling strategy was proposed as a first solution for handling larger temporal displacements in MCTF. The novel content-adaptive MCTF enables the proposed method to adapt to different motion types in each sequence. The experimental results show that the learned MCTF video coder exhibits promising rate-distortion performance, especially for higher bitrates. On the UVG data set, the MCTF-CA method achieves average BD rate savings of -21\% and -9\% for GOP sizes of 4 and 8, respectively, compared to HM. Thereby, it clearly outperforms the state-of-the-art video coder DCVC-HEM \cite{Li2022}.

There are various possibilities for improvement as an initial version of a learned wavelet video coder. First, one could examine a different temporal subband coder required for practical usage because the autoregressive context model of iWave++ prohibits parallelization. Second, the MCTF structure requires extensions to handle more diverse motion types and GOP sizes of 16 and higher. Because the maximum frame distance doubles with every additional temporal decomposition level, motion estimation is considerably more challenging for, for example, a GOP size of 16 with a frame distance of 8. Therefore, bidirectional motion estimation and methods for overcoming the limitations of short-sequence training sets for larger GOP-size compression could be investigated. To mitigate ghosting for larger GOP sizes, an adaptive choice of a truncated DWT without temporal update \cite{Turaga2005} could be beneficial. Furthermore, the complexity of content-adaptive MCTF can be limited by using a predictor for choosing the adaptive MCTF option.

The MCTF-based approach provides an explainable and scalable alternative to common autoencoder-based video coders. This paper made the first steps to enable further development of this important direction of research.

\bibliographystyle{IEEEtran}
\bibliography{wavelets}

\end{document}